\documentclass[12pt,epsf]{article}
\usepackage{epsfig}

\newcommand{\onefigure}[2]{\begin{figure}[htbp]
\begin{center}\leavevmode\epsfbox{#1.eps}\end{center}\caption{#2\label{#1}}
\end{figure}}
\newcommand{\setfigure}[2]{\begin{figure}[htbp]
\begin{center}\leavevmode\epsfxsize=5in\epsfbox{#1.eps}\end{center}\caption{#2\label{#1}}
\end{figure}}
\newcommand{\twofigures}[3]{\begin{figure}[htdp]
\centering \leavevmode\epsfxsize=2.5in\epsfbox{#1.eps}
\leavevmode\epsfxsize=2.5in\epsfbox{#2.eps} 
\caption{{
#3}\label{#1}}
\end{figure}}
\newcommand{\threefigures}[4]{\begin{figure}[htdp]
\centering \leavevmode\epsfxsize=2in\epsfbox{#1.eps}
\leavevmode\epsfxsize=2in\epsfbox{#2.eps}
\leavevmode\epsfxsize=2in\epsfbox{#3.eps} \caption{\small {\em
#4}\label{#1}}
\end{figure}}

\renewcommand{\thanks}[1]{\footnote{#1}} 

\newcommand{\be}{\begin{equation}}
\newcommand{\ee}{\end{equation}}
\newcommand{\bea}{\begin{eqnarray}}
\newcommand{\eea}{\end{eqnarray}}

\begin{document}

\pagestyle{empty}

\bigskip\bigskip
\begin{center}
{\bf \large Quantum Behaviors on an Excreting Black Hole}
\end{center}

\begin{center}
James Lindesay\footnote{e-mail address, jlslac@slac.stanford.edu} \\
Computational Physics Laboratory \\
Howard University,
Washington, D.C. 20059 
\end{center}
\bigskip

\begin{center}
{\bf Abstract}
\end{center}

Often, geometries with horizons
offer insights into the intricate
relationships between general relativity and quantum physics.
However, some subtle aspects of gravitating quantum systems
might be difficult to ascertain using static backgrounds, since
quantum mechanics incorporates dynamic measurability
constraints (such as the uncertainty principle, etc.).

For this reason, the behaviors of quantum systems on a dynamic
black hole background are explored in this paper. 
The velocities and
trajectories of representative outgoing, ingoing, and
stationary classical particles are calculated and
contrasted, and the dynamics
of simple quantum fields (both massless and massive)
on the space-time are examined. 
Invariant densities associated with the quantum fields are
exhibited on the Penrose diagram that represents the 
excreting black hole.

Furthermore,
a generic approach for the consistent mutual gravitation
of quanta in a manner
that reproduces the given geometry is developed. 
The dynamics of the mutually gravitating quantum fields
are expressed in terms of the affine parameter that describes
local motions of a given quantum type on the space-time. 
Algebraic equations that relate the energy-momentum
densities of the quantum fields to Einstein's tensor
can then be developed.  An example mutually gravitating system
of macroscopically coherent quanta along with a core gravitating
field is demonstrated.  
Since the approach is generic and
algebraic, it can be used to represent a variety of
systems with specified boundary conditions.

\bigskip \bigskip \bigskip

\setcounter{equation}{0}
\section{Introduction}
\indent

There have been remarkably few direct experimental
observations of the quantum behaviors of gravitating
systems.  Experiments by Overhauser, et.al.\cite{Overhauser},
have demonstrated that those dynamic gravitational
fields local to the Earth's surface do not break the coherence
of gravitating neutrons, which give interference results
consistent with the principle of equivalence.  
Those experimental results involve both Newton's
gravitational constant $G_N$ and Planck's constant
$\hbar$ in a single equation form.  Such
results, along with other observed phenomena,
require that gravitating sub-clusters can maintain quantum
coherence while having their internal dynamics influenced
by dis-entangled clusters co-contributing to the local
gravity.  Although justifications of the redshift of gravitating
photons need not make use of Planck's constant, those
quanta do maintain their coherence during extended
interactions with dynamic gravitational fields (e.g.
the cosmic microwave background, local gravitational
redshift measurements, etc.).  Similarly,
interacting gravitating systems continue to gravitate
after dis-entanglement, which motivates a formulation
that incorporates straightforward cluster decomposability
within macroscopic gravitational environments.

Generally, the dis-entanglement of relativistic dynamic
quantum clusters that is necessary for correspondence with
classical dynamics requires that the geometric aspects
of the kinematics associated with the Lorentz /Poincare
transformation properties of a given cluster
must be separate and distinct from the internal
coherent descriptions and off-shell analytic behaviors
of disparate clusters. 
The methods utilized in this paper incorporate the
techniques developed in the establishment of cluster
decomposable formulations in relativistic few particle
scattering\cite{LMNP}.  Those formulations exhibit
the expected dis-entanglement of interacting quantum
scattering states needed for
classical correspondence properties, in
spite of the kinematic complications introduced by
the non-linear nature of the relativistic energy-momentum
dispersion relation for massive systems.  The solution
requires proper cluster independence of the geometric
parameters describing the kinematics between subsystems
from the internal quantum dynamics associated with
the description of an interacting system in terms of
the boundary (i.e., only \emph{self}-interacting)
states.  Cluster-decomposable
relativistic scattering theory is most directly realized
as follows:
\begin{itemize}
\item the clusters should be characterized using the channel
decomposition that Faddeev\cite{Fadeev} developed for describing
cluster-decomposable unitary non-relativistic systems;
\item the kinematics between external and intermediate
quantum states (\emph{off-diagonal} dynamics) should insure
Lorentz frame (3-velocity)
conservation rather than 3-momentum conservation.  Velocity
conservation has been referred to as the \emph{point form} 
representation of the dynamics
by Dirac\cite{Dirac} and others, whereas momentum
conservation has been referred to as the
\emph{contact form} representation;
\item the internal dynamics of the intermediate quantum
states (or descriptions of the interacting state in terms of
the complete basis of boundary states) for the various
non-interacting clusters should be independent;
\item the complex analytic extension of the dynamic
invariant energy of a given cluster (the \emph{off-shell} behavior)
should only parametrically affect the kinematics of the
other clusters, i.e., the internal dynamics of one cluster
should not alter the energy spectrum of another.
\end{itemize}
Formulations of scattering theory with these properties
have been shown to have appropriate
non-relativistic behaviors\cite{AKLN}
as well as to give the expected results from perturbative
representations such as quantum electro-dynamics\cite{Compton}.

The relevant quantum dynamics
is most conveniently expressed in
terms of the affine parameter associated with the gravitating
particles/fields. 
The affine or proper time derivative of local physical parameters
can be expressed in terms of the substantive derivative
${d \over d \lambda}=u^\beta {\partial \over \partial x^\beta }
\equiv u^\beta \partial_\beta $. 
This then makes the
incorporation of the principles of
equivalence/relativity straightforward,
resulting in substantive quantum
flows on the geometry\cite{Harry}.  

Macroscopic quantum fluids (e.g. superfluids and superconductors)
maintain persistent quantum flows that satisfy
quantization conditions in the non-inertial
environments of most laboratory measurements. 
For instance, vortices of superflow with quantized circulation
maintain the angular momentum of a rotating vessel of liquid helium
cooled below the superfluid transition temperature. 
(It would be quite interesting to see if the precession of vortices
with quantized circulation ${h \over m}$ in a
gravitational field $G_N$ is a
measurable phenomenon.) 
Fluid continuity directly follows from the equations of
motion resulting from Lagrangians constructed using
substantive derivatives.  Superfluid flows then result from
the local gauge invariance of those Lagrangian forms. 
Thus, substantive flows have been quite useful
in describing macroscopic behaviors of quantum fluids.

Substantive quantum flows inherently articulate the
philosophical approach of proper time formulations\cite{Gill}
that presume the primacy of coordinates associated with the
system being influenced in descriptions of the interaction.
One expects that
the affects of external interactions upon a given system
can be most fundamentally understood in terms of the dynamical
parameters of that impacted system. 
Thus, when combined with the principle of equivalence, the
dynamics of free quantum fields in a gravitating
environment can be most elegantly described using
the affine (proper) parameters of that quantum field.
Furthermore,
formulations that exemplify cluster decomposibility and
substantive affine flows are most
directly realized by developing linear spinor fields
consistent with those in reference \cite{LinearSpinor}.  The
form of the gravitating quanta developed in
Section \ref{Section4} will involve affine flows of independent
states motivated by the previous discussions.

The general incorporation of quantum mechanics into
gravitating formulations requires that the descriptions
be dynamic, since the relationships between temporal
and kinematic parameters have inherent constraints
due to the quantum measurability problem (e.g. the
uncetainty principle, non-commuting measurables, etc.). 
Therefore, the background geometry will be taken to be
that of a dynamic excreting
spherically symmetric space-time with a metric form given by 
\be
ds^2 = -\left (1-{R_M (ct) \over r} \right ) (dct) ^2 +
2 \sqrt{{R_M (ct) \over r}} dct \, \, dr + dr^2 + r^2 \, 
(d\theta^2 + sin^2 \theta \, d\phi^2).
\label{metric}
\ee
In this equation, $R_M (ct) \equiv 2 G_N M(ct) / c^2$ is a time-dependent
form of the Schwarzschild radius that
will be referred to as the
\emph{radial mass scale}.   For this dynamic black hole, the
radial mass scale is not a light-like surface, and is not coincident
with the horizon (which \emph{is} a light-like surface). 
However, a traversing outgoing light-like trajectory is momentarily
at rest in the radial coordinate $r$ as the radial mass scale
moves past that trajectory. 
The trajectory of the horizon is determined
by calculating the null radially outgoing
geodesic $ds^2=0$ that intersects
the radial mass scale as the mass of the black hole vanishes. 
This metric was developed as a dynamic
extension of the river model of (static) black holes discussed in the
literature\cite{rivermodel}.  The asymptotic ($r
\rightarrow \infty $) form of the metric
is that of a Minkowski space-time.  Therefore, the coordinates
$(ct,r,\theta,\phi)$ are the temporal, radial, and angular
coordinates of an asymptotic observer.  

The temporal parameter $t$ used to describe the dynamics
of the black hole metric \ref{metric}
is not singular at the horizon.  More generally, physical curvatures
are non-singular\cite{JLMay07} away from $r=0$. 
The contravariant form for the Einstein tensor calculated
from this metric can be expressed in terms of the
excretion rate $\dot{R}_M(ct)$ and the dimensionless parameter
$\zeta \equiv {R_M (ct) \over r}$, taking the form
\be
((G^{\mu \nu})) \,=\,
\left (
\begin{array}{c c c c}
0 & 0 & 0 & 0 \\
0 & -{\dot{R}_M (ct) \over r^2 \sqrt{\zeta}} & 0 & 0 \\
0 & 0 & -{\dot{R}_M (ct) \over 4 r^4 \sqrt{\zeta}} & 0 \\
0 & 0 & 0 & -csc^2 (\theta){\dot{R}_M (ct) \over 4 r^4 \sqrt{\zeta}}
\end{array}
\right ) .
\label{EinsteinTensor}
\ee
Since several components of the Einstein tensor
are non-vanishing, this dynamic geometry provides an excellent
laboratory for the exploration of self-consistent
gravitating phenomena.

The subsequent developments will exhibit the behaviors of
classical and quantum objects on the black hole background
described in Eqns. \ref{metric} and \ref{EinsteinTensor}. 
Section \ref{Section2} will exhibit solutions to the geodesic
equations for the geometry, as well as demonstrate typical
trajectories of massive outgoing, ingoing, and stationary
classical particles on the Penrose diagram of the space-time. 
The expected behaviors of classical massless particles
have been confirmed in a previous paper\cite{BABJLphotons}.
The behavior of a spherically symmetric Klein-Gordon field
will be exhibited in Section \ref{Section3}.  The
form of the Klein-Gordon
field will be found to be clumsy for constructing mutually
gravitating fields, so a more convenient set of cluster-decomposable
scalar fields will be developed in Section \ref{Section4}. 
Finally, in Section \ref{Section5} an example mutually
gravitating macroscopic system of quanta is exhibited and discussed.

\setcounter{equation}{0}
\section{Geodesic Motion
\label{Section2}}

\subsection{Four velocities}
\indent

Geodesic motion is at the foundation of the principle
of equivalence.  The geodesic equation describes
the evolution of four-velocities on the geometry. 
In all calculations, four-velocities of both massive 
(${dx^\beta \over d c \tau}$)
and massless (${dx^\beta \over d \lambda}$) particles
will be taken to be dimensionless.  The form
satisfied by the radial component
of four-velocities on the metric \ref{metric}
is given by
\be
u^r = - \zeta^{1/2} \, u^0 \pm \sqrt{(u^0)^2-\Theta_m}
\label{urEqn}
\ee
where 
$\Theta_m \equiv 
\left \{
\begin{array}{l}
0 \quad m=0 \\
1 \quad m \not= 0
\end{array} \right .  .$
In particular, stationary massive particles satisfy
$u^0=1, u^r=-\zeta^{1/2}$.
Those four-velocities associated with
geodesic motion satisfy
the geodesic equations:
\be
{d u^0 \over d \lambda}=
-{\zeta^{1/2} \over 2 r} (\zeta^{1/2} u^0 + u^r)^2 ,
\label{u0Eqn}
\ee
and
\be
{d u^r \over d \lambda}=
-{\dot{R}_M \over 2 r \zeta^{1/2}} (u^0)^2 - \Theta_m {\zeta \over 2 r}.
\label{urEqn}
\ee

Away from the horizon $\zeta=\zeta_H$ or the ingoing
horizon $\zeta=\zeta_I$ (which of course define the geodesic
of a particular outgoing or ingoing massless particle,
respectively), the four-velocities of freely gravitating particles
can be reparameterized in terms of the dimensionless
variable $\zeta$ in the black hole geometry.  This
reparameterization involves the transformation
\be
{d \zeta \over d \lambda}={\dot{R}_M u^0 - \zeta u^r \over r} .
\ee
Generally, affine derivatives can be expressed in terms of the
substantive (or \emph{flow}) derivative
${d \over d \lambda} = u^\beta \partial_\beta$,
which, using Eqns. \ref{u0Eqn} and \ref{urEqn}
implies that gravitating four-velocities can be
expressed as functions of the single parameter $\zeta$, i.e.
$u^\beta=u^\beta (\zeta)$.

Therefore, numerical solutions of the geodesic equations
can be directly obtained as ordinary differential equations.
For all numerical calculations, the (dimensionless) excretion rate will
be taken to be constant $\dot{R}_M=0.1$.  This produces an
outgoing horizon dressing the singularity $r=0$ at
$\zeta_H \cong 1.177$ and an ingoing horizon
of outermost communication with the singularity at
$\zeta_I \cong 0.078$.  Plots of the four-velocities of
massless particles with asymptotic values $u^0(0)=1$
are demonstrated in Figure \ref{M0plu0ur}, while
plots of the four-velocities of massive particles with
asymptotic values $u^0(0)=1.2$ are demonstrated in
Figure \ref{Mplu0ur}.
\twofigures{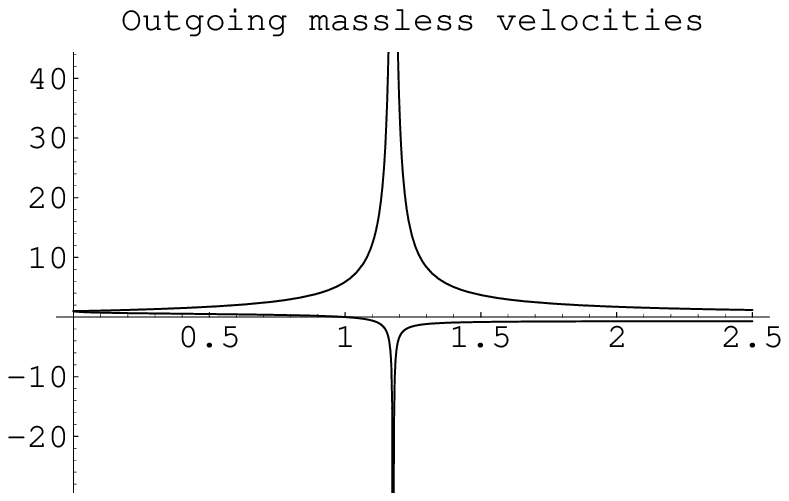}{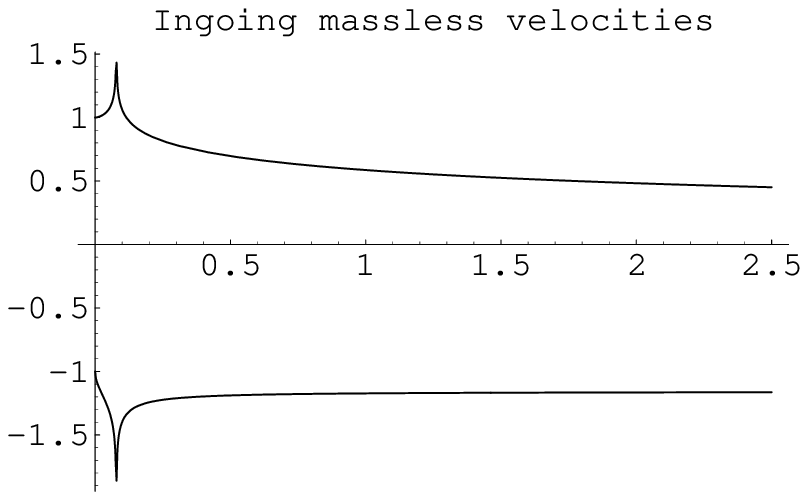}{Four velocities for massless
particles as a function of $\zeta$.  Upper curves represent $u^0$ and lower $u^r$}
\twofigures{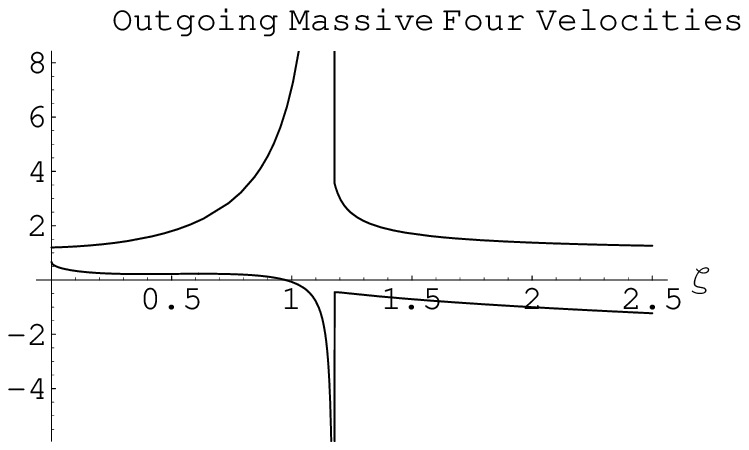}{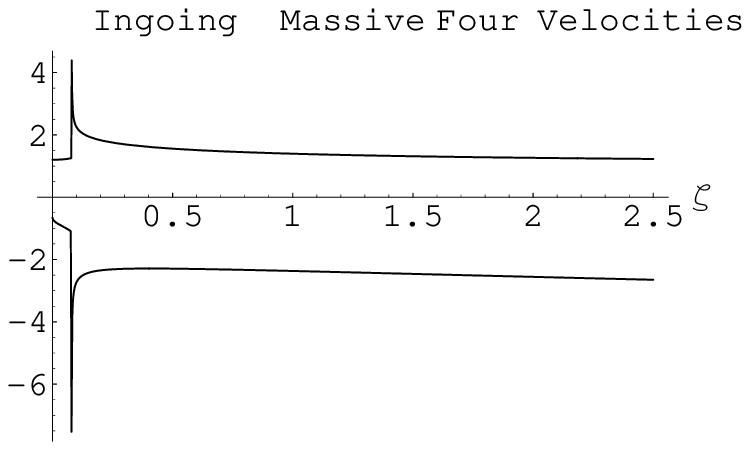}{Four velocities for massive
particles as a function of $\zeta$.  Upper curves represent $u^0$ and lower $u^r$}
The four-velocities of outgoing particles are seen to have
singular behavior at the horizon $r=R_H (ct)$,
while the ingoing particles have
singular behavior at the ingoing horizon $r=R_I (ct)$.  The radial
components of the four-velocity of outgoing
particles are seen to change sign at the radial mass scale
$\zeta_M =1$. Stationary
massive particles are described by a vanishing value for the
radical in Eqn. \ref{urEqn}.  The plot of stationary particles in the space-time
is demonstrated in Figure \ref{Su0ur}.
\onefigure{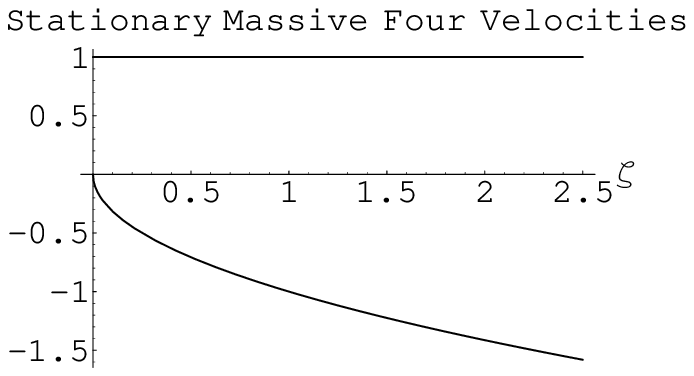}{Stationary massive $u^0$ (upper plot) and $u^r$
(lower plot).}
Any of the above-calculated forms for the four-velocities
of massless and massive systems 
can then be used to develop the classical
trajectories of gravitating particles.

\subsection{Form of conformal coordinates}
\indent

Space-time diagrams are always useful in the
visualization of dynamics in a given
geometry.  In particular, Penrose diagrams are
convenient for examining the large-scale causal
structure of a geometry because of the following
properties:
\begin{list}{...}{}
\item Penrose diagrams map the entire space-time
onto a single finite page, and
\item Penrose diagrams preserve the slope of
light-like trajectories at $\pm$unity.
\end{list}
Since light-like curves are easy to plot on a Penrose
diagram, the causal structure of the geometry
can be directly observed from the diagram, and
potential causal relationships between locations
can be immediately ascertained.  In order to
construct the Penrose diagram for a given space-time,
conformal coordinates, which give unity slope for
light-like curves, must be found.  For Minkowski
space-time, the coordinates $(ct,r)$ are already
conformal.  However, for the metric \ref{metric},
conformal coordinates must be constructed, since
clearly null geodesics will not have unity slope for
those coordinates.

For a black hole satisfying Eqn. \ref{metric} with a constant
rate of mass accretion/excretion $\ddot{R}_M=0$, the
conformal coordinates can be determined in a straightforward
manner.  The forms of the conformal temporal and radial coordinates
that are used for the construction of the Penrose diagram
have been developed in a companion paper\cite{BABJL1},
and are given by
\be
\begin{array}{c}
ct_* = {r \over 2} \left (
 exp \left [ \int ^ {R_M (ct) \over r} { \left (
1 + \sqrt{\zeta '}  \right ) d \zeta '  \over
\left \{ \zeta' \left ( 1 + \sqrt{\zeta'} \right )
+ \dot{R}_M \right \} }
\right ]  -  
 exp \left [ \int ^ {R_M (ct) \over r} { \left (
1 - \sqrt{\zeta '}  \right ) d \zeta '  \over
\left \{ \zeta' \left ( 1 - \sqrt{\zeta'} \right )
- \dot{R}_M \right \} }
\right ] \right ) \\ \\
r_* = {r \over 2} \left (
 exp \left [ \int ^ {R_M (ct) \over r} { \left (
1 + \sqrt{\zeta '}  \right ) d \zeta '  \over
\left \{ \zeta' \left ( 1 + \sqrt{\zeta'} \right )
+ \dot{R}_M \right \} }
\right ]  + 
 exp \left [ \int ^ {R_M (ct) \over r} { \left (
1 - \sqrt{\zeta '}  \right ) d \zeta '  \over
\left \{ \zeta' \left ( 1 - \sqrt{\zeta'} \right )
- \dot{R}_M \right \} }
\right ] \right )  .
\end{array}
\label{conformal}
\ee
This equation relates the space-time coordinates of an
asymptotic observer $(ct, r)$ with the conformal coordinates
$(ct_* , r_*)$.  The Penrose diagram of a 
spatially coherent black hole that excretes at a constant
rate $\dot{R}_M$ until it vanishes at $t=0$, is shown in
Figure \ref{EvapPenr}.
\setfigure{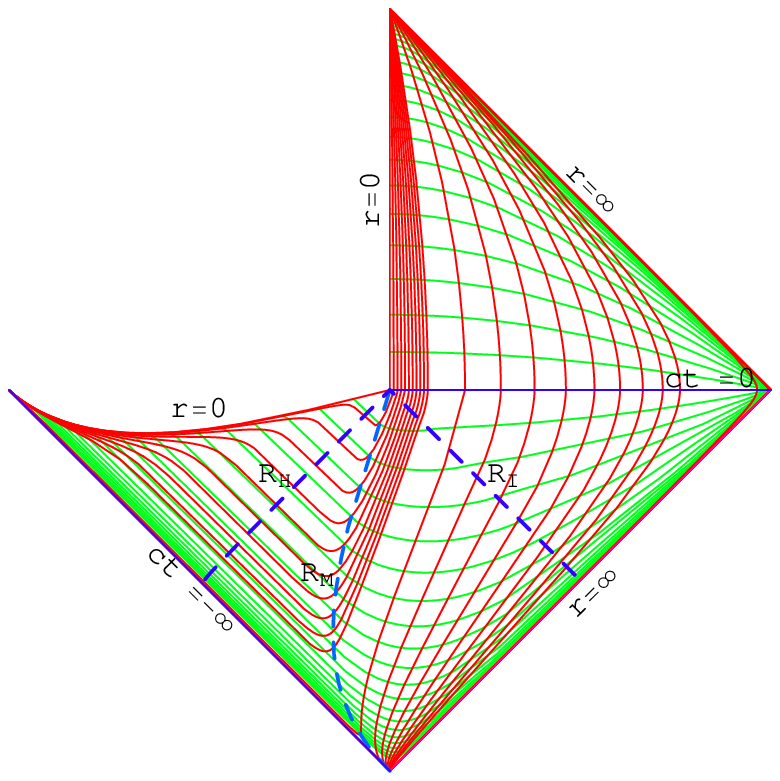}{Penrose diagram for a black hole that evaporates steadily
to zero mass at $ct=0$.  Red curves (running vertically in the right
hand region) represent curves of constant $r$ successively 
in tenths of units, units of length, and decades of units.  The green curves
represent curves of constant $ct$ in units of length. 
The horizontal solid blue line on the right
represents the end of excretion $ct=0$ $(\zeta_0=0)$, while the
horizontal red curve on the left
represents the space-like singularity of
the black hole $r=0$ $(\zeta_s = \infty )$.
The dashed blue line labeled
$R_H$ $(\zeta_H \approx 1.177)$ represents the horizon,
that labeled $R_M$ $(\zeta_M = 1)$ represents
the radial mass scale, and that labeled $R_I$
$(\zeta_I \approx  0.078)$ represents the ingoing horizon.}
The singularity of the black hole is represented by the bounding
space-like curve $r=0$ on the left-hand side of the diagram.  At
$t=0$, this singularity vanishes, and the curve $r=0$ becomes
time-like in the final Minkowski space-time (which is represented by the
upper right quadrant of the diagram).  No communication to the
left of the light-like horizon $R_H$ can escape hitting the
singularity.  Likewise, no communication to the right of the
light-like incoming horizon $R_I$ can communicate with the
singularity.  The radial mass scale $R_M(ct)=2 G_N M(ct) / c^2$
is seen to itself be a time-like trajectory where outgoing light-like
trajectories remain temporarily stationary in coordinate $r$.  The
conformal coordinates $(ct_*,r_*)$ in Eqn. \ref{conformal}
are chosen to correspond to
the Minkowski coordinates $(ct,r)$ (which are, of course, also conformal)
of the asymptotic observers far from the black hole. 
Therefore, the bounding light-like curves on the right-hand quadrants
of the diagram correspond to the Minkowski space-time of distant
observers.

\subsection{Classical trajectories}
\indent

The classical trajectories of freely falling bodies can be directly
represented on the Penrose diagram by integrating the
four-velocities subject to the appropriate initial conditions.
All light-like trajectories are indeed found to always have slopes
of $\pm$unity\cite{BABJLphotons}.  Massive particle trajectories
are demonstrated in Figure \ref{Mcurves}.
\twofigures{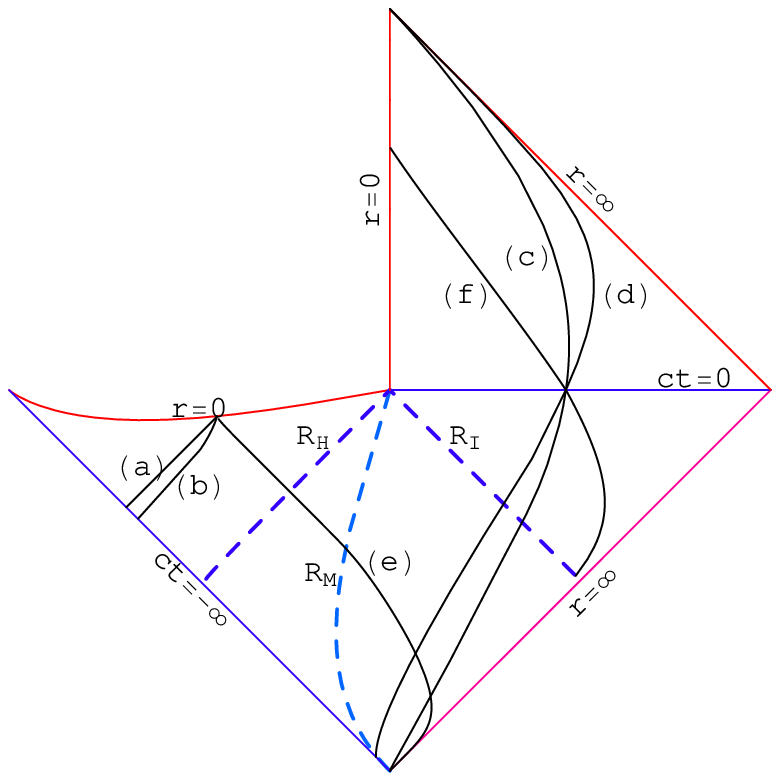}{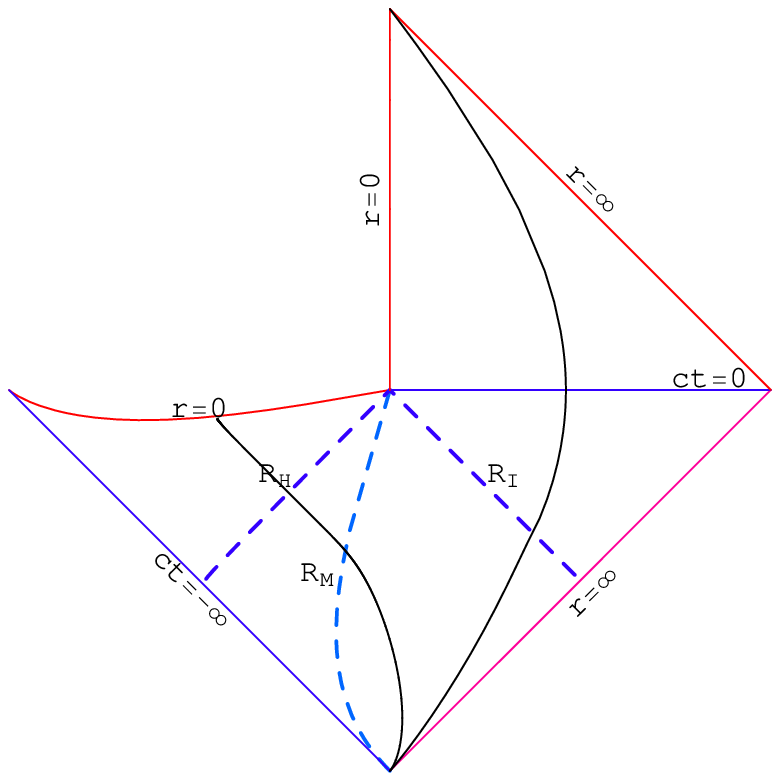}{Classical trajectories for massive
gravitating particles.  The Penrose diagram on the left demonstrates
outgoing trajectories (a) and (b) terminating on the singularity
at $ct=-5$ , outgoing trajectories (c) and (d) passing through
$r=+5$ at $ct=0$, and ingoing trajectories (e) terminating at $ct=-5$, and (f) passing
through $r=+5$ at $ct=0$.  The diagram on the right demonstrates stationary
trajectories terminating on the singularity at $ct=-5$, and passing through
$r=+5$ at $ct=0$.  }
In the first diagram, various outgoing and ingoing trajectories
are plotted.  Trajectory (a) is an outgoing mass that ultimately
hits the singularity, while trajectory (b) is an outgoing
mass that hits the singularity at the same time,
but moving more slowly.  Trajectory
(c) is an outgoing trajectory that escapes hitting the singularity,
while trajectory (d) is a faster outgoing trajectory that passes
through the point $(ct=0,r=5)$ at the same time.  Trajectory
(e) is an ingoing mass that hits the singularity at the same
time as trajectories (a) and (b).  Trajectory (f) is an ingoing mass
that is external to the ingoing horizon $R_I$, and passes
through the point $(ct=0,r=5)$ at the same time as trajectories
(c) and (d).  No numerically stable solution for massive outgoing
trajectories in the region between the horizon and the
radial mass scale was found. 
The second diagram plots the trajectories of
stationary masses.  The trajectory on the left of the second diagram
represents a stationary mass that hits the singularity at the
same time as trajectories (a), (b), and (e) on the first diagram. 
The trajectory on the right of the second diagram represents a
stationary mass that passes through the point $(ct=0,r=5)$
at the same time as trajectories (c), (d), and (f) on the first diagram. 
All trajectories that escape the singularity are seen to
smoothly transition from gravitating projectiles
to inertial free motion as the space-time transitions from
dynamic black hole to Minkowski space-time at $t=0$. 
It is interesting to note that a stationary particle
which remains in the exterior
region of the black hole geometry apparently
follows the same trajectory
using conformal coordinates as it would in Minkowski space-time.

\setcounter{equation}{0}
\section{Klein-Gordon Field
\label{Section3}}
\indent

The primary focus of this paper is the examination of quantum
behaviors on the dynamic space-time defined by the
metric \ref{metric}.  It is instructive to first examine the
quantum mechanics of a well-understood scalar field
on the space-time background described by this metric.  The
Lagrangian of a non-interacting gravitating massless scalar
Klein-Gordon field $\chi$ will take the form
\be
\mathcal{L}={1 \over 2} \sqrt{-g} \, \,  g^{\mu \nu}
\partial_\mu \chi^*(x) \, \partial_\nu \chi (x) .
\ee
The massless field is chosen so that no additional scale
is introduced into the problem by including a mass for the quanta. 
As shown in reference \cite{LSJLBlackHoles}, one can perform
a partial wave expansion on $\chi$
\be
\chi (x) \equiv \sum_{\ell m}{\psi_\ell (ct,r) \over r} \, Y_\ell ^m (\theta, \phi)
\ee
to obtain equations\cite{JLNSBP08}
describing the dynamics of the field $\psi_\ell$:
\be
\begin{array}{r}
-{\partial^2 \psi_\ell \over (\partial ct)^2}
+{\partial \over \partial ct} \left [
\sqrt{R_M \over r} \left (  {\partial \psi_\ell \over \partial r} \right ) \right ]
+{\partial \over \partial r} \left [
\sqrt{R_M \over r} \left (  {\partial \psi_\ell \over \partial ct} \right ) \right ] 
+{\partial \over \partial r} \left [
\left (1-{R_M \over r} \right ) \left (  {\partial \psi_\ell \over \partial r} \right ) \right ] + \\ \\
- \left [{\ell (\ell + 1) + {R_M \over r} \over r^2} + {1 \over r} {\partial \over \partial ct}
\left ( \sqrt{R_M \over r} \right )
\right ] \psi_\ell=0 .
\end{array}
\label{KGEulerLagrange}
\ee

S-wave ($\ell=0$) solutions of Eqn. \ref{KGEulerLagrange}
have been numerically calculated for the excreting black hole. 
Boundary conditions at $ct=0$ require that the field $\psi_0$
and its derivatives must appropriately match
those of a free-space Minkowski
Klein-Gordon field near that region. The matching
Minkowski space solution for
the field away from $r=ct$ is given by
\be
\psi_M (ct,r) =\left [ A +B \, log
\left ( {r+ct \over r-ct} \right ) \right ]
\Theta (r-ct) .
\ee
Plots of s-wave solutions to Eqn. \ref{KGEulerLagrange} on the black
hole background are shown in Figure \ref{KGPsi}.
\twofigures{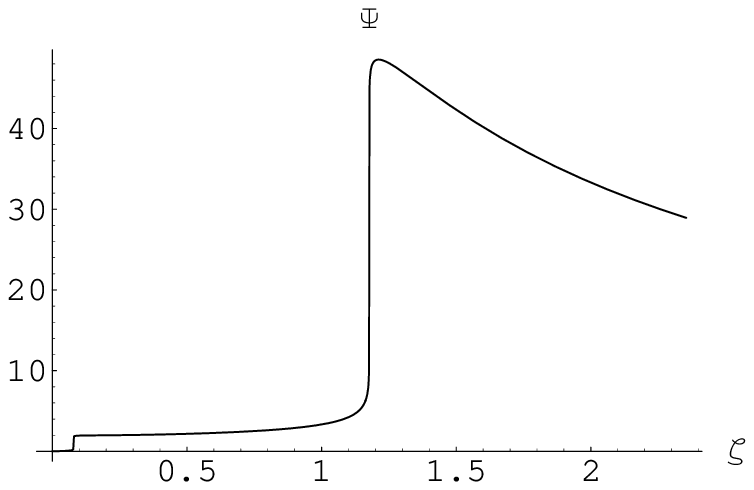}{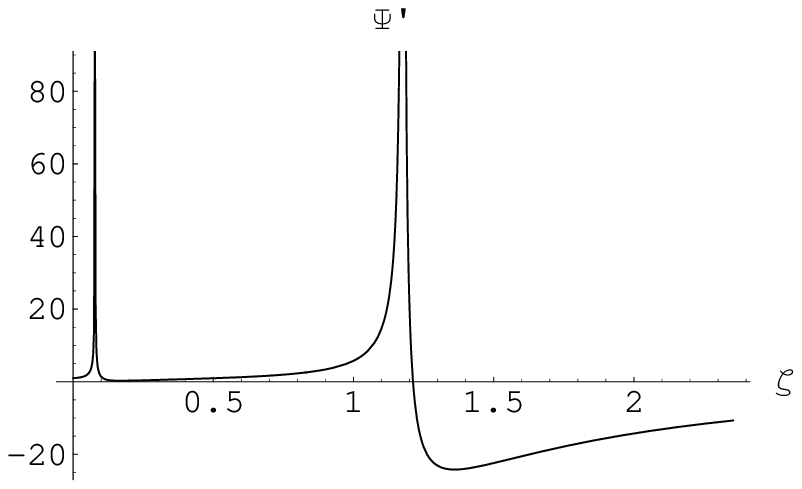}{Plots of $\psi(\zeta)$ and
$\psi'(\zeta)$ for a massless Klein-Gordon field on
the dynamic black hole
background, where $\zeta=R_M (ct)/r$.  Field normalization units
are arbitrary.}
The field was chosen to have a very small (though non-vanishing)
value asymptotically ($\zeta=0$).  The wavefunction is continuous
throughout the space-time, but its derivative is 
seen to have singular behavior near
the ingoing horizon $R_I$ $(\zeta_I \approx  0.078)$ and the horizon
$R_H$ $(\zeta_H \approx 1.177)$.  The magnitude of the
scalar field is seen to be largest within and near the horizon.

To gain physical insight into the distribution of the scalar
field on the global space-time, a density plot on the Penrose
diagram is instructive.  For generating such a plot, 
the computer was instructed to place a pixel at a point on the 
Penrose diagram corresponding to the asymptotic observer's 
coordinate label $(ct, r)$ if the density multiplied by a random 
number bounded by 0 and 1 was larger than the value calculated at a 
normalization point.  The plot then demonstrates relative measures 
of the density throughout the space-time.  The quantity
plotted in Figure \ref{KGProbDens} is the probability density
$|\chi (ct,r)|^2$. 
\twofigures{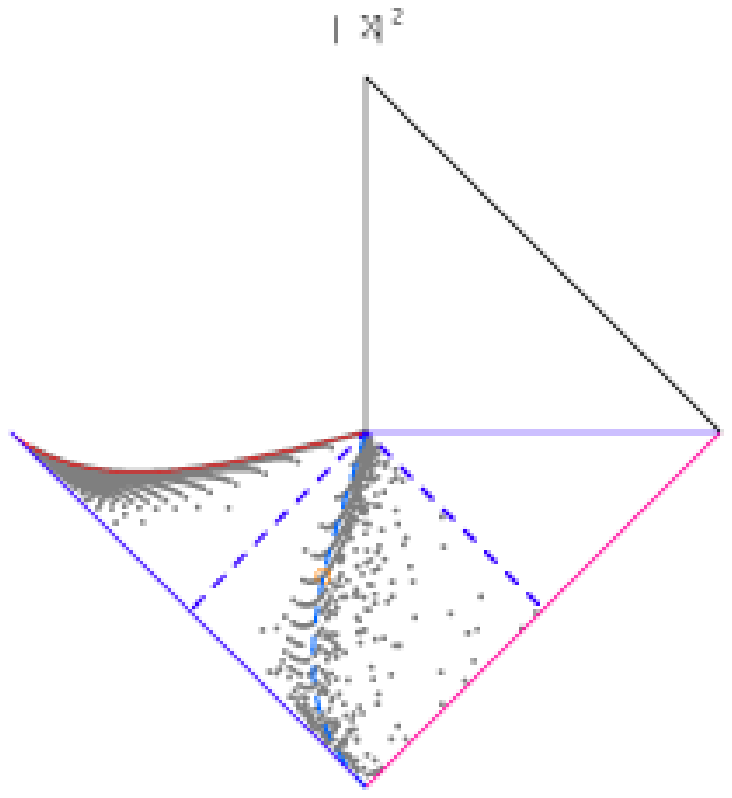}{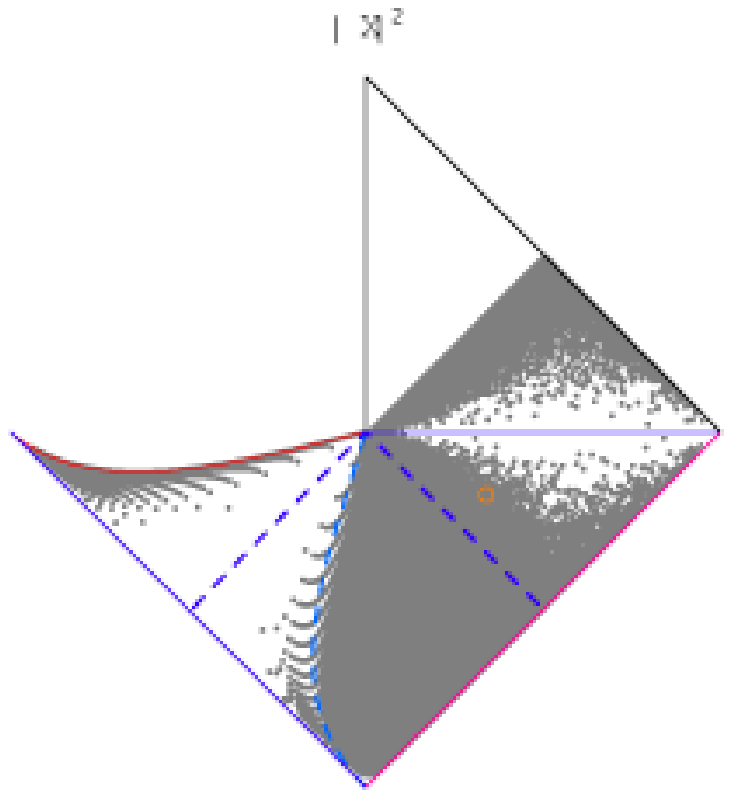}{Penrose density plot
of a spherically symmetric gravitating massless Klein Gordon field
(left) and that same field's correspondence with the
Minkowski space-time solution (right).}
In the first density plot, the normalization point (represented
by the center of the small circle)
was placed on the radial mass scale curve $R_M (ct=-5)$. 
The second curve represents the same density, only placing
the normalization point just outside of the inner horizon, which
allows smaller values of the field density to be plotted.  The
weak field black hole solution for the field is seen to reflect
the Minkowski space-time solution that matches that field
on the line (volume) $ct=0$.

An intriguing question is whether the spherically
symmetric massless Klein-Gordon
field can self-gravitate to produce this black hole geometry.  The
energy-momentum tensor generated by the Lagrangian
\ref{KGEulerLagrange} can be calculated several ways.  For
this field, the action was calculated from
the Lagrangian using $W=\int \mathcal{L} \, d^4 x$ . 
The energy-momentum tensor was then calculated by
examining the behavior of this action under variations of the metric,
$\delta W \equiv {1 \over 2} \int d^4 x \sqrt{-g} \, T^{\mu \nu} \,
\delta g_{\mu \nu}$.  One can then compare this
tensor to the Einstein tensor \ref{EinsteinTensor} to examine
how the scalar field
might contribute to the gravitating system.  In particular,
it is convenient to examine the
physical scalar density invariant
$g_{\mu \nu} T^{\mu \nu}$ (i.e. the trace of the scalar field's
energy-momentum tensor), which would be directly related to the
Ricci scalar $\mathcal{R}$ of the space-time if the Klein-Gordon
field was self-gravitating.  The Klein-Gordon density invariant
and the Ricci scalar are plotted in Figure \ref{KGInvar}.
\twofigures{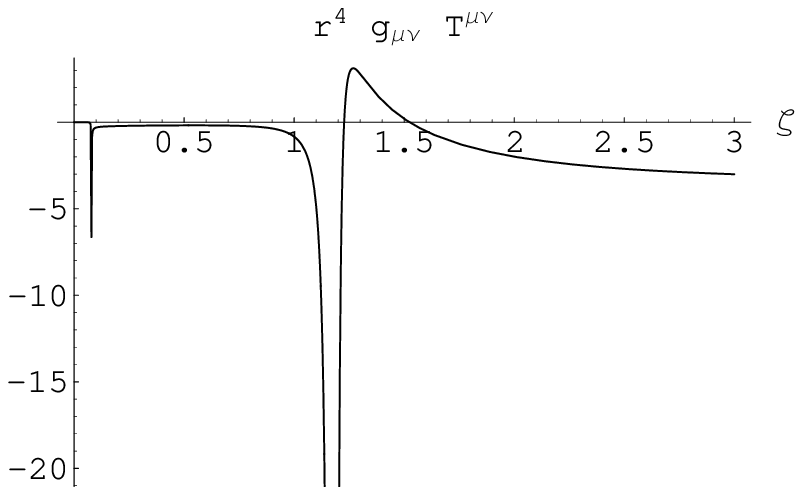}{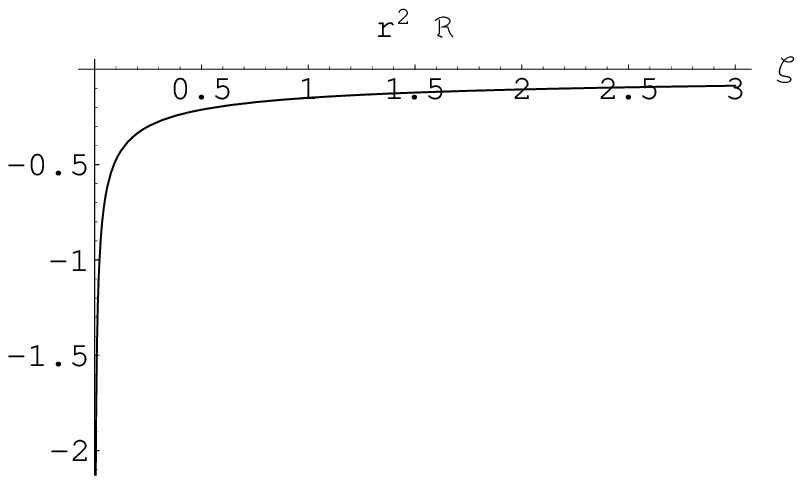}{Plots of invariant scalar densities
associated with gravitating quanta on the black hole background.  The
left diagram plots $r^4$ times the trace of the energy-momentum
tensor of the Klein-Gordon field.  The diagram on the right plots $r^2$
times the Ricci scalar of the dynamic black hole.  The horizontal
axis for both diagrams is the dimensionless parameter $\zeta=R_M(ct)/r$.}
The density invariant is singular at the horizon and ingoing
horizon, while the Ricci scalar is not. 
The Klein-Gordon field here calculated was not expected to
directly construct a self-gravitating quantum consistent solution
to Einstein's equation.  The non-linear form of the Klein-Gordon
energy-momentum tensor in derivatives
makes superpositions clumsy.  That form also makes the
dependency of the density invariant upon the radial coordinate
$r$ irreparably different from that of the Ricci scalar. 
In addition, the form does not
intuitively incorporate some of the properties that one
would expect of a self-gravitating field.  In the next section,
a foundation will be developed for the exploration of
gravitating quantum fields on a space-time
background that can be superposed in a manner that might
construct mutually gravitating objects.

\setcounter{equation}{0}
\section{Cluster-Decomposable Gravitating Scalar Fields
\label{Section4}}
\indent

Useful gravitating quanta that mutually contribute to a
particular space-time configuration are expected to satisfy
certain physical properties:
\begin{list}{...}{}
\item it should be possible to have dis-entangled components
that gravitate, and can co-contribute to a mutually gravitating
environment;
\item the energy densities of the components should linearly
contribute to the overall energy density used to drive the
Einstein equation for the gravitating environment;
\item there should be components that can maintain quantum
coherence, satisfying probability flux conservation and
expected energy/time-momentum/position phase relationships;
\item gravitating quantum systems should be able to maintain
observed linearity as previously dis-entangled components
interact, forming changing configurations of those systems;
\item the collective gravitational field should incorporate
the principle of equivalence with regards to the gravitation of
a given component, i.e., one should obtain expected
behaviors when performing
scattering and spectroscopic experiments in a locally
inertial freely falling frame.
\end{list}

Interacting few-particle quantum systems exhibit peculiar
behaviors, such as the Efimov effect\cite{Efimov} (which demonstrates
long range coherence for short ranged interactions), even
for weakly interacting non-relativistic systems. 
The non-linear behavior of the energy-momentum relationship
in Minkowski space-time further complicates attempts to
explicitly demonstrate how interacting quantum systems can
dis-entangle.  The problem is that the complicated relativistic
kinematics between entangled clusters will generally alter
descriptions of the dynamic properties of one cluster when another
cluster is interacting.  For instance, in naive relativistic formulations
an electron interacting on the moon would 
substantially affect the energy spectrum
of a hydrogen atom here on earth. 
However, cluster-decomposable formulations of quantum scattering
maintain relativistic covariance, and incorporate correspondence
with the classical dis-entanglement of clusters\cite{LMNP, AKLN}.
The flat space-time solution of cluster-decomposable relativistic
quanta requires that one properly separate the geometry
(kinematics) from the off-shell quantum dynamics.  As was
mentioned in the Introduction, this is
done by formulating the dynamics while incorporating
the following properties:
\begin{list}{}{}
\item (i)  a recognition of the difference between off-shell
and off-diagonal descriptions of intermediate states.  The
off-shell behavior describes the analytic structure of the
invariant energy description of the scattering
amplitudes in the complex plane.  Bound
states manifest as poles in the off-shell parameter.  The
off-diagonal behavior describes how the fully interacting
system can be described in terms of a complete set of boundary
states. The intermediate (virtual) state dynamics is separate
from the off-energy-shell parametric behavior;
\item (ii)  the off-diagonal description of intermediate quantum states
should use Lorentz frame conservation (parameterized by preserving
the three components of the four-velocity $\mathbf{u}$, or
so called \emph{point form} dynamics)
instead of momentum $\mathbf{p}$ conservation (or \emph{contact
form} dynamics);
\item (iii)  the formulation should
use parametric (geometric rather than quantum
dynamic) descriptors of cluster kinematics, in terms of the
invariant energies $\vec{u} \cdot \vec{P}$ for the overall
system, and $\vec{u}_a \cdot \vec{P}_a$ for cluster $a$.
\end{list}
In order to most directly infuse these properties into the
formulation, quantum states with explicit  linear behaviors
in derivatives will be constructed.

\subsection{Dynamical equations of gravitating quanta}
\indent

As previously mentioned, Lagrangians that utilize
substantive derivatives\cite{Harry} most directly incorporate
the spirit of the principle of equivalence.  The chosen form
for the Lagrangian of a gravitating non-interacting cluster $a$
in the geometry will be taken to be
\be
\mathcal{L}_a \equiv \sqrt{-g} \,\, \textnormal{L}_a =
-\sqrt{-g} \,\, [ {i \hbar c \over 2} u^\beta
(\psi_a^* \partial_\beta \psi_a -( \partial_\beta \psi_a^*)  \psi_a)
-m_a c^2 \psi_a^* \psi_a ] .
\label{RadiationLagrangian}
\ee
Such a Lagrangian form can be directly generalized to
incorporate linear spinor fields\cite{LinearSpinor}.

Writing the functions $\psi_a$ as complex
parameters $\psi_a \equiv |\psi_a| e^{i \xi_a}$,
the Euler-Lagrange equations then insure
probability conservation
\be
{1 \over \sqrt{-g}} \partial_\beta (\sqrt{-g} \,
|\psi_a|^2 u^\beta)=0 ,
\label{ProbabilityConservation}
\ee
and phase coherence
\be
u^\beta \partial_\beta \xi_a = - {m_a c \over \hbar} .
\label{PhaseCoherence}
\ee
Therefore, systems described by the Lagrangian form
Eqn. \ref{RadiationLagrangian} satisfy the dynamics expected
of inertial quantum fields. 
Substituting the phase coherence equation \ref{PhaseCoherence}
back into
the original form for the Lagrangian, it can be shown that
the extremum Lagrangian for the gravitating quanta
vanishes $\mathcal{L}_a =0$.

If one defines $E_m$ as the proper value (standard form
invariant of the little group\cite{WeinbergQFT}) of the energy of
the particle (i.e. $mc^2$ for massive particles, $E_0$ for massless
particles),  a general solution to Eqn. \ref{PhaseCoherence} is
given by
\be
\partial_0 \xi = {E_m \over \hbar c} (u_0 + Q \, u^r )=
{E_m \over \hbar c} \left [
-(1-\zeta)u^0 + (\sqrt{\zeta}+Q) u^r 
\right  ] ,
\label{d0XiEqn}
\ee
\be
\partial_r \xi = {E_m \over \hbar c} (u_r - Q \, u^0)=
{E_m \over \hbar c} \left [
(\sqrt{\zeta}-Q)u^0 + u^r
\right  ] ,
\ee
where the covariant forms of the 4-velocities are given by
\be
\begin{array}{l}
u_0 = (\zeta - 1) u^0 + \zeta^{1/2} u^r , \\ \\
u_r = \zeta^{1/2} u^0 + u^r .
\end{array}
\ee
The four-velocities satisfy the usual normalization
$u^\beta u_\beta=-\Theta_m$.

The integrability of a function of several variables
means that the function is in fact well described
in terms of those variables, satisfying the analytic property
of equality of second partial derivatives regardless of order. 
The integrability of the phase $[\partial_r , \partial_0] \,
 \xi (ct,r) =0$ implies
that the function $Q$ must satisfy
\be
u^\beta \partial_\beta Q = \partial_0 u_r
- \partial_r u_0 - (\partial_\beta u^\beta) \, Q.
\label{QEqn}
\ee
Since the four-velocities depend only on the
the dimensionless parameter $\zeta$, 
$u^\beta=u^\beta({R_M (ct) \over r})$, 
the functions $Q=Q(\zeta)$ can likewise be shown to
depend only upon the dimensionless parameter $\zeta$.
For the metric form \ref{metric}, derivatives of the
geodesic four-velocities (from Eqns. \ref{u0Eqn}
and \ref{urEqn}) satisfy
\be
\partial_\beta \, u^\beta = {\Theta_m \over r} {\zeta^{1/2} \over 2 }
{\dot{R}_M + \zeta^{3/2} \over \dot{R}_M u^0 - \zeta u^r },
\label{uDivergence}
\ee
and
\be
\partial_0 u_r-\partial_r u_0=0 .
\label{uCurl}
\ee

Numerical solutions for the magnitudes and phases
of the fields can be developed by defining the
dimensionless parameters $|\tilde{\psi}(\zeta)|^2$ by the equation
$|\psi(ct,r)|^2 \equiv {|\tilde{\psi}(\zeta)|^2 \over L_P \, r^2}$,
and the reduced phases $\tilde{\xi}(\zeta)$ (which carry
the dimension of inverse Compton wavelength) by the 
equation $\xi(ct,r) \equiv r \tilde{\xi}(\zeta)$.  It is interesting
to note that $|\tilde{\psi}(\zeta)|^2$ and $Q(\zeta)$ have
the same form for their differential equations.

For massless particles, one can immediately determine
that the parameters $|\tilde{\psi}(\zeta)|^2$ and
$Q(\zeta)$ are constants.  Analytic solutions
can also be obtained for stationary particles.  For
stationary massive particles, the solutions take the form
\be
{|\tilde{\psi}_S(\zeta)|^2 \over |\tilde{\psi}_S(\zeta_o)|^2}=
\left ( {\dot{R}_M + \zeta_o ^{3/2} \over
\dot{R}_M + \zeta ^{3/2} } \right )^{1/3}=
{Q_S(\zeta) \over Q_S(\zeta_o)} .
\ee
Numerical forms describing the quantum dynamics of
outgoing and ingoing massive quanta can directly be
obtained by solving the straightforward ordinary differential
equations for the phase factor $Q(\zeta)$ (as well as the
reduced squared wavefunction $|\tilde{\psi}(\zeta)|^2$)
using Eqn. \ref{QEqn}, and integrating Eqn. \ref{d0XiEqn} to obtain
the phase $\xi$.

\subsection{Energy-momentum tensors of gravitating quanta}
\indent

The form of the contribution of the gravitating quanta 
to the energy-momentum tensor
can be directly calculated using the
standard Legendre transformation from the Lagrangian form
to the Hamiltonian form of the dynamics of a
multi-component field $\chi$:  
\be
{\mathcal{T}^\beta} _\mu= \sum {\partial \mathcal{L} \over
\partial (\partial_\beta \chi)} \partial_\mu \chi  -
\delta_\mu^\beta \mathcal{L} .
\ee
The Euler-Lagrange equations imply that
the extremal form of the Lagrangian \ref{RadiationLagrangian}
vanishes $\mathcal{L}_a=0$ for any quantum  $a$.  Therefore,
the energy-momentum tensor of quantum $a$ satisfies
\be
{\mathcal{T}_a^\beta} _\mu=\sqrt{-g} {T_a^\beta} _\mu=
\sqrt{-g} (\hbar c) u_a^\beta (\partial_\mu \xi_a) |\psi_a|^2 .
\label{SubstRadTbetamu}
\ee
The energy-momentum contribution of quantum $a$
to the overall
system is seen to depend linearly upon derivatives of
the phase, directly linking the local phase of the wavefunction
$a$ to its contribution to the local energy-momentum of the
system.

The form of these generic quanta need not locally conserve
energy-momentum due to the background gravitation.  The
generic form for the divergence of the of the energy-momentum
tensor of a field $\psi_a$ is given by
\be
{T_a^\beta}_{\mu \, ;\beta} = 
- {\Gamma_{\mu} ^\lambda}_\beta {\partial \textnormal{L}_a
\over \partial ({\partial_\beta \psi_a})} \partial_\lambda \psi_a +
complex \, conjugate.
\ee
Neither is the form of this tensor necessarily
symmetric, satisfying
\be
T_a^{\, 0 \, r}-T_a^{\, r \, 0}=  \Theta_{m_a}
{m_a c \over \hbar} \, |\psi_a|^2 \, Q_a .
\ee
Since the Einstein tensor \ref{EinsteinTensor} is both
geometrically conserved and symmetric, the collective
form of the energy-momentum tensor driving the
Einstein equation must likewise be locally
conserved and symmetric.

\setcounter{equation}{0}
\section{Macroscopic Self-Gravitating Quanta
\label{Section5}}
\indent

Although the form of microscopic quantum gravity
remains uncertain, the quantum mechanics of
systems co-gravitating with macroscopic
media has been phenomenologically explored by common
experience as well as experiment\cite{Overhauser}.  
As a complement to attempts to self-consistently describe the 
microscopic behaviors between mutually gravitating
quanta, one should be able to gain insight into the
fundamentals of gravitation by developing micro-physical
behaviors that consistently reproduce an example
dynamic space-time.  
Therefore
the form of mutually gravitating coherent sub-clusters
that are consistent with the metric Eqn. \ref{metric} will
be developed in this section.

The geometrodynamics of the space-time must 
consistently co-mingle with the micro-physical behaviors
for the construction to be viable.  The Ricci scalar
$\mathcal{R}$ used in the calculation of the Einstein
tensor is directly related to a physical invariant given
by the trace of the energy-momentum tensor ${T^\beta}_\beta$.
For the space-time metric \ref{metric}, it is given by
\be
\mathcal{R}={3 \dot{R}_M \over 2 r^2} \sqrt{r \over R_M} ,
\label{RicciScalar}
\ee
which is non-singular away from a physical singularity
at the origin, and vanishes for a static radial mass
scale $\dot{R}_M\rightarrow 0$ for finite radial mass
scales $R_M$.  
However, one should note that although
the metric Eqn. \ref{metric} takes the same form as a Minkowski
space-time as the radial mass scale vanishes,
some physical attributes of
a black hole that excretes at a steady
rate become singular when $R_M \rightarrow 0$
in Eqn. \ref{RicciScalar}.  This
paper is not concerned with how the excretion rate $\dot{R}_M$
turns off as the radial mass scale vanishes, which is what must
occur in an actual physical process. 
Reasonable calculations show that
one can obtain viable quantum solutions with stepwise constant rate
equations for radial mass scales down to the Planck length.  Therefore,
gravitating quantum solutions will be examined for the constantly
excreting space-time represented in the lower quadrants of the
Penrose diagram in Figure \ref{EvapPenr} for times $t<0$.

\subsection{Core quantum field}
\indent

The Lagrangian densities for the
cluster-decomposable gravitating scalars developed
in Section \ref{Section4} vanishes for extremal fields.
The space-time described by the metric \ref{metric} is
spherically symmetric, so that all velocities have vanishing
angular components $u^\theta =0= u^\phi$.  This means
that quanta with
energy-momentum tensors of the form
Eqn. \ref{SubstRadTbetamu} cannot contribute to
the non-vanishing components of the Einstein
tensor ${G^\theta}_\theta$ and ${G^\phi}_\phi$. 
Therefore, the vanishing of the Lagrangian densities
$\mathcal{L}_a =0$ for the non-interacting
gravitating quanta, requires the addition of
an interacting form for a core
gravitating field in order 
to be consistent with Einstein's equation.

The core gravitating field is expected to relate closely
to classical geometric parameters, and will be represented
by real fields with a Lagrangian of the form
\be
\begin{array}{r}
\mathcal{L}_{core}=-\sqrt{-g} \, \hbar c \left [
\psi_{c+} (u^\beta \partial_\beta \psi_{c+} + {1 \over 2}
u^\beta \partial_\beta (log \sqrt{-g}) \psi_{c+}) \right . + \\
-\left . \psi_{c-} (u^\beta \partial_\beta \psi_{c-} + {1 \over 2}
u^\beta \partial_\beta (log \sqrt{-g}) \psi_{c-})
\right ] .
\end{array}
\label{CoreLagrangian}
\ee 
The core fields $\psi_{c\pm}$ satisfy the same Euler
Lagrange equations, but will have opposing signs for
their energy densities.
The core gravitating fields are seen to have an interaction
term dependent on the local form of the metric.

For the real fields $\psi_{c \pm}$, the Euler-Lagrange
equations require that
\be
\sqrt{-g} \, \psi_{c \pm} \, \partial_\beta u^\beta =0.
\ee
From Eqn. \ref{uDivergence}, one observes that \emph{any}
massless core fields will satisfy these Euler-Lagrange
equations.  Light-like propagation properties for the
core gravitating field are, of course, intuitively gratifying. 
This gives considerable flexibility in choosing
forms for the core fields $\psi_{c \pm}$ that
are consistent with the requirements of Einstein's
equation for this geometry. 

The form of the core field energy-momentum tensor
is given by
\be
{{T_{core}}^\beta}_\mu = -\hbar c \left [
 \psi_{c+} u^\beta \partial_\mu  \psi_{c+}
-  \psi_{c-} u^\beta \partial_\mu  \psi_{c-}
\right ] - \delta_\mu^\beta \mathcal{L}_{core}/\sqrt{-g} .
\label{CoreTbetamu}
\ee
Einstein's equation then requires that the physical content
of the system satisfies
\be
{G^\beta}_\mu=-8 \pi {L_P^2 \over \hbar c} ( {{T_{core}}^\beta}_\mu +
{{T_{rad}}^\beta}_\mu  )
\label{EinsteinEqn}
\ee
where the Planck length is defined by
$L_P \equiv {\hbar \over M_p c} =
\sqrt{\hbar G_N \over c^3}$ and the tensor of radiating
quanta $T_{rad}$ is the collective form of the cluster-decomposable
quanta developed in Section \ref{Section4}.  
As previously stated, since the radiations
defined in Eqn. \ref{RadiationLagrangian} cannot contribute
to ${G^2}_2 = {G^3}_3$ for a spherically symmetric
system (since $u^\theta = 0 = u^\phi$ in the equation
\ref{SubstRadTbetamu} for the radiating components'
contributions), Einstein's equation constrains
the form of the core gravitating field from Eqn. \ref{EinsteinTensor},
\be
-{\dot{R}_M \over 4 r^2 \sqrt{\zeta}}= {G^3}_3 = 
-8 \pi {L_P^2 \over \hbar c} {{T_{core}}^3}_3 = 
8 \pi {L_P^2 \over \hbar c} \mathcal{L}_{core} / \sqrt{-g} ,
\ee
which yields the equation satisfied by the core gravitating fields
\be
{\dot{R}_M \over 16 \pi L_P^2 \, r^2 \sqrt{\zeta}}=
( u^\beta \partial_\beta  \psi_{c+}^2 + {2 u^r \over r}  \psi_{c+}^2 )
-( u^\beta \partial_\beta  \psi_{c-}^2 + {2 u^r \over r}  \psi_{c-}^2 ) .
\label{CoreFieldEqn}
\ee

A solution of Eqn. \ref{CoreFieldEqn}
with positive semi-definite probability
densities can be found.  The solution
required (i) outgoing massless core quanta, (ii) $\psi_{c+}=0$
for $r<R_H(ct)$ and (iii) $\psi_{c-}=0$
for $r>R_H(ct)$.  If described in terms of a single
core field, the square of that single field simply changes
signs across the horizon. 
Solutions are demonstrated for reduced core
probability density $|\tilde{Y_c}(\zeta)|^2$ (where
$|\psi_c(ct,r)|^2 \equiv {|\tilde{Y}_c(\zeta)|^2 \over L_P^2 \, r}$)
and reduced core energy density ${L_P^2 \, r^2 \over \hbar c}
T_{core}^{0 \, 0} $
in Figure \ref{PsiCore}.
\twofigures{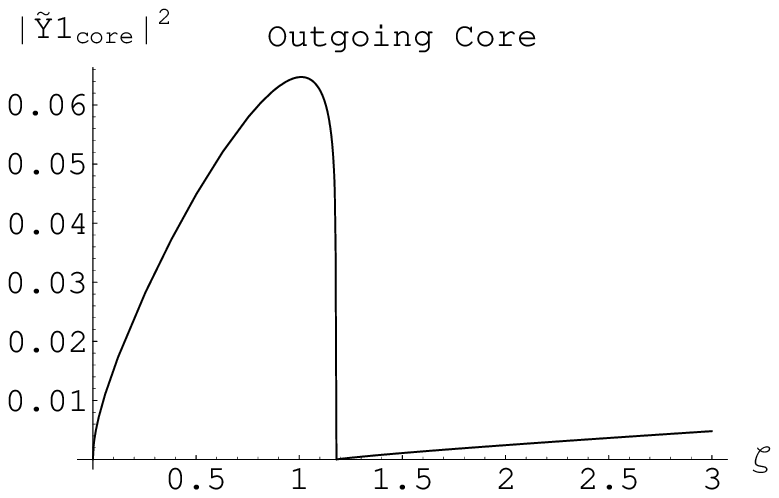}{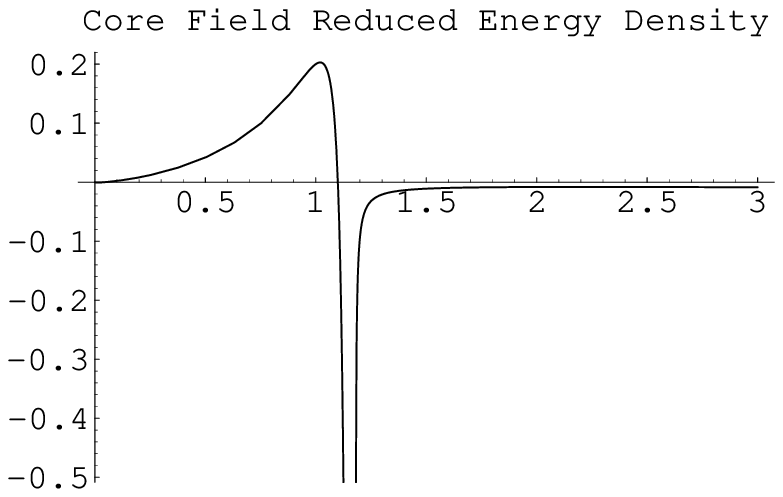}{Reduced wavefunction squared and energy
density of core field as a function of $\zeta=R_M (ct)/r$.}
The solution for the square of the core wavefunction is everywhere
non-negative, but vanishes at the horizon $\zeta_H \cong 1.177$. 
The core field energy density is negative in regions distant from the
black hole (for $\zeta <  0.2$), positive in the region outside but
near the horizon $\zeta_H$, and again negative inside the
horizon.  Since the component $G^{00}$ vanishes, the sign taken
by the energy density of the radiating quanta from Section \ref{Section4}
must cancel the energy density contributed by the core field. 
It was therefore crucial that the asymptotic form of the core
field have this negative energy density so that normal gravitating
particles far from the black hole have the expected sign for their
energy densities.

As was seen for the Klein-Gordon field in Section \ref{Section3}
it is illustrative to examine a density plot of the physical densities
calculated in Figure \ref{PsiCore} on the global space-time
represented by the Penrose diagram.  As was previously discussed,
only that portion of the Penrose diagram satisfying $t<0$
will be physically populated. 
The square of the
core field wavefunction and energy density are plotted in
Figure \ref{ProbCore}.
\twofigures{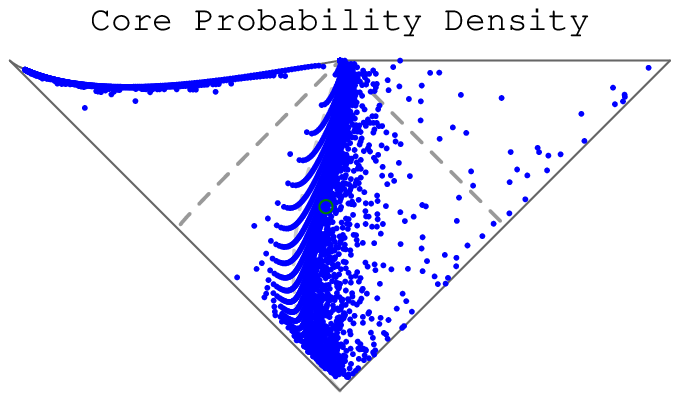}{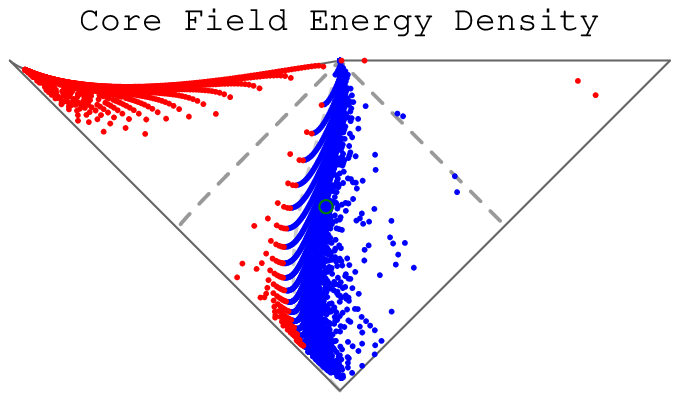}{Penrose density plots 
of core field probability
density and energy density.}
As with all density plots, these plot represent the relative
magnitude of the given density in terms of the relative number
of pixel points plotted in a given vicinity.  The normalization
point is represent by the small circle in the middle of the diagram
near the radial mass scale $\zeta_M=1$.  Since the physical
parameters being plotted might take on negative values, a blue
pixel is plotted if the density is positive relative to the
absolute value of the normalization scale,
while a red pixel is plotted if
it is negative relative to the negative absolute value of the
normalization scale.  The core field energy density is seen
to take on negative values in the interior region, as well as
small negative values in far regions on the right of the Penrose
diagram.

Solutions of Eqn. \ref{CoreFieldEqn} involving ingoing
massless core quanta were likewise calculated. 
The solution for ingoing massless core quanta obtained
by the author had negative probability densities in
a small region
near the radial mass scale (within numerical accuracy), and
were not further explored.

\subsection{A mutually-gravitating macroscopic system}
\indent

Once the core gravitating field of the excreting black hole has been
determined from Eqn. \ref{CoreFieldEqn}, the linear behavior of
the radiation fields can be exploited to construct a 
mutually/self gravitating system.
Substituting solutions for the core field calculated from
Eqn. \ref{CoreFieldEqn} into
the core energy-momentum in Eqn. \ref{CoreTbetamu}, this
energy-momentum tensor then placed into Einstein's
equation \ref{EinsteinEqn} defines a form that must be satisfied
by the collection of radiating gravitating quanta.

From Eqn. \ref{SubstRadTbetamu}, a generic form
for the macroscopic gravitating quanta is given by
\be
{{T_{rad}}^\beta}_\mu=\sum_a 
(\hbar c) u_a ^\beta (\partial_\mu \xi_a) |\psi_a|^2 .
\label{RadiationTbetamu}
\ee
It is very gratifying that in the exterior region of the black hole,
there are many, many solutions satisfying Einstein's equation
\ref{EinsteinEqn} due to the algebraic and
dis-entangled nature of the 
contributions from the radiating quanta.
This means that the phenomenologically observed behaviors of
dynamic spherically symmetric stars, planets, etc. can 
likely be
incorporated using mutually gravitating particles as described
by the linearly additive Eqn. \ref{RadiationTbetamu}.

An example solution for a mutually gravitating
macroscopically coherent
system of quanta will be directly demonstrated.  Since there are 4
components of the Einstein tensor Eqn. \ref{EinsteinTensor}
that remain to be satisfied by Eqn. \ref{EinsteinEqn},
a minimal algebraic solution should be obtained using
4 particle types in Eqn.  \ref{RadiationTbetamu}.  The geometry
should have coherent emissions/absorptions of otherwise
conserved radiating
quanta locally contributing to the energy-momentum of the 
system in the form
\be
\begin{array}{c}
{{T_{rad}}^\beta}_\mu =   (\hbar c)  \left [
u_1 ^\beta (\partial_\mu \xi_1) N_1 |\psi_1|^2 +
u_2 ^\beta (\partial_\mu \xi_2) N_2 |\psi_2|^2 + \right . \\ \\ \left .
u_3 ^\beta (\partial_\mu \xi_3) N_3 |\psi_3|^2 +
u_4 ^\beta (\partial_\mu \xi_4) N_4 |\psi_4|^2
\right ] ,
\end{array}
\label{CoherentTrad}
\ee
where $N_a$ represents the number of quanta of type $a$ present
in the gravitating system.  As long as the fields are linearly
independent, solutions to Einstein's equation \ref{EinsteinEqn} using
macroscopic quantum fields of the form in Eqn. \ref{CoherentTrad} can
always be found.  However, one is not guaranteed that physical boundary
conditions with positive semi-definite number densities can be
found.  It was found to be quite straightforward to obtain physically
meaningful solutions with positive number densities and positive
mass quanta in the region exterior to the radial mass scale $r>R_M$. 
However, the author has yet to develop solutions that have both
positive semi-definite number densities and positive semi-definite
masses in all regions.  This is likely due to the behavior
of the component $u^r$ as seen in Eqn. \ref{urEqn},
since it can change its sign in the
interior region of the black hole, and therefore affect the
algebraic solutions. 
An example macroscopic mutually
gravitating excreting black hole is given below.

The solution demonstrated in Figures \ref{Xi1}, \ref{Xi3},
\ref{NumDens}, and \ref{Prob1Pen} involve a system with
four particle types.  Particle types 1 and 2 are stationary
gravitating quanta with mass $m$ and linearly independent
phases demonstrated in Fig. \ref{Xi1}.  
\threefigures{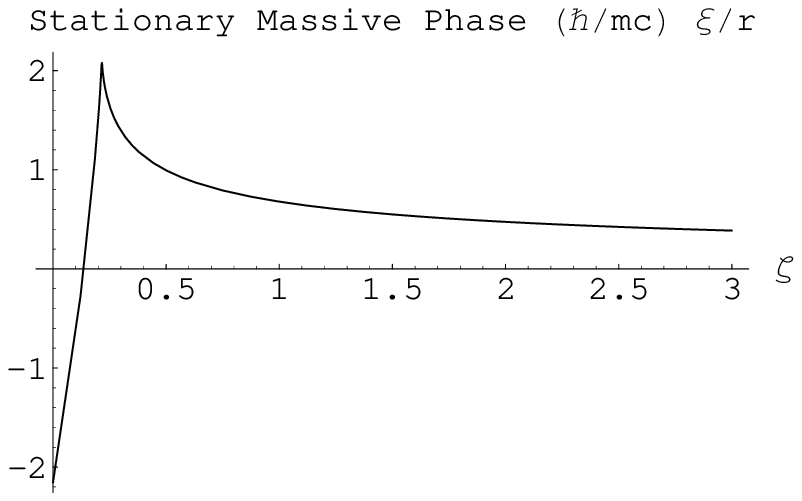}{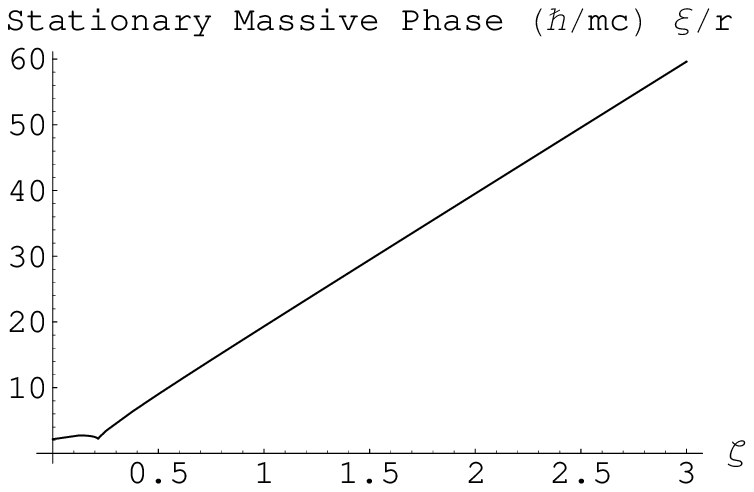}{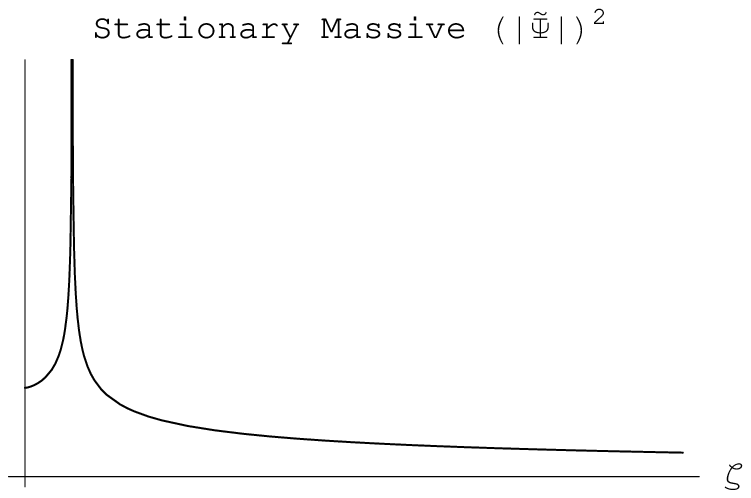}{Phases of stationary mass 1 and mass 2,
and the square of the reduced stationary wavefunction.}
The reduced phases $\tilde{\xi} (\zeta) \equiv
{\hbar \over m c} {\xi(ct,r) \over r}$ of the stationary particles
are represented in the first two diagrams in the figure. 
The magnitude squared of the reduced wavefunction
of the stationary particles $|\tilde{\psi}_a(\zeta)|^2 \equiv
r^2 L_P |\psi_a(ct,r)|^2$ is represented in the third diagram in the figure.
The singular behavior in the stationary form of $|\tilde{\psi}(\zeta)|^2$
occurs at $\zeta_{SS}=(-\dot{R}_M)^{2/3} \cong 0.215$ for the excretion
rate chosen for this paper, which is in the region of the space-time
between the radial mass scale $R_M$ ($\zeta_M=1$) and the ingoing horizon
$R_I$ ($\zeta_I \cong 0.078$).   In Figure \ref{Xi3}
particle type 3 is an outgoing massless quantum represented
by the first diagram and particle type 4 is an ingoing massless quantum
represented by the second diagram. 
\twofigures{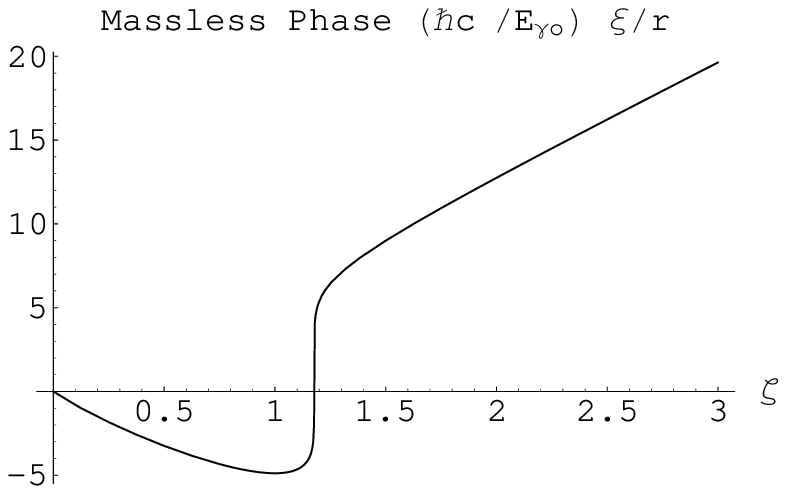}{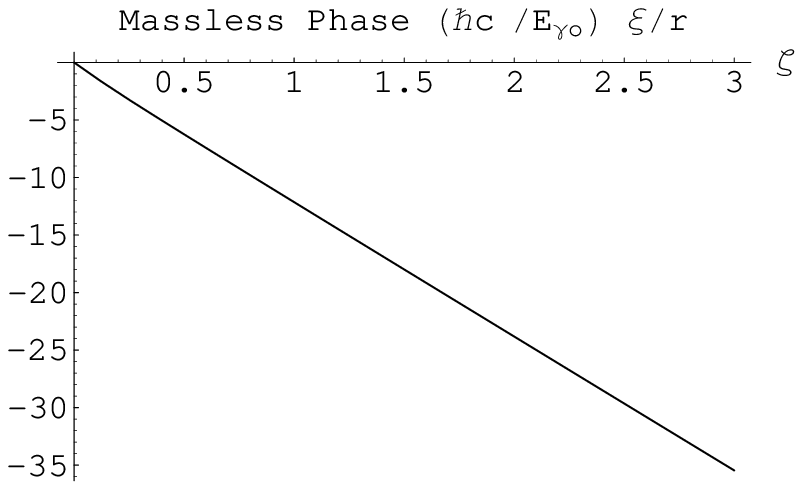}{Outgoing and ingoing massless particle phases. }
The reduced
wavefunction for massless quanta is constant, and the reduced
phase is expressed in terms of the 
quantum's asymptotic energy $E_o$ using $\tilde{\xi} (\zeta) \equiv
{\hbar c \over E_o} {\xi (ct,r) \over r}$.

Algebraic solutions for the reduced number densities of the four
particle types $\mathcal{N}_a (\zeta) \equiv 
{E_{m_a }\over \hbar c} L_P^2 \,
r^2 \, N_a (ct,r) |\psi_a (ct,r)|^2$ (where again $E_m$
is the \emph{proper} or \emph{affine}
energy form of a quantum of mass $m$), are demonstrated
in Figure \ref{NumDens}.
\setfigure{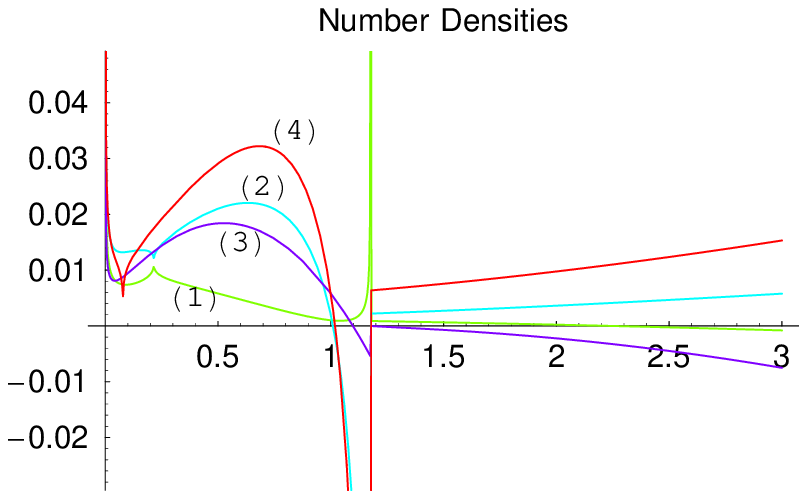}{Reduced number densities of mutually gravitating
quanta.  Particle types (1) and (2) are stationary massive
quanta.  Particle type (3) is an outgoing massless quantum,
while particle type (4) is an ingoing massless quantum.  The
horizontal axis is the dimensionless scale $\zeta=R_M (ct) / r$}
The plot representing a particular particle type is directly labeled
on the diagram.  The affine energy form $E_{m_a}$
(associated with the affine parameter describing the particle motion)
appears in the phase of the wavefunction of particle type $a$
as is the case with standard quantum mechanics.  For massive
states it represents the ``co-moving" energy $m c^2$, while
for massless states is represents the ``standard state"
energy-momentum $E_0$ of the particle observed in
the boundary inertial reference frame. 
One can see that particle types 2, 3, and 4
must take on negative values of  $E_m$
in the region just inside of the radial mass scale $r=R_M (ct) \,
(\zeta_M =1 )$, and outside of the horizon $r=R_H (ct)  \,
(\zeta_H \sim 1.2)$,
if the quantum number densities of those particle types are to
remain positive semi-definite.  Thus, one expects to find some
negative mass stationary quanta of type 2 
just inside the radial mass scale, but outside of the horizon.  Particle type 1 is seen to require a
negative mass value inside the horizon near the singularity.
Since the solutions to Eqn. \ref{EinsteinEqn} are algebraic and only depend upon
Einstein's equation, the number densities themselves are independent of the normalization
of the wavefunctions.  Only the actual numbers of quanta satisfying
a particular flux normalization depend
upon the form of that normalization.  Since the problem examined
here has unbounded past temporal extent, normalization properties
will be examined using models for a complete finite duration
life-cycle (accretion, then evaporation) of a
black hole in subsequent papers.

It is particularly illustrative to examine density plots of the
various number densities on the Penrose diagram.  Density
plots of the square of the wavefunction and
number density of particle type 1 are displayed in Figure \ref{Prob1Pen}.
\twofigures{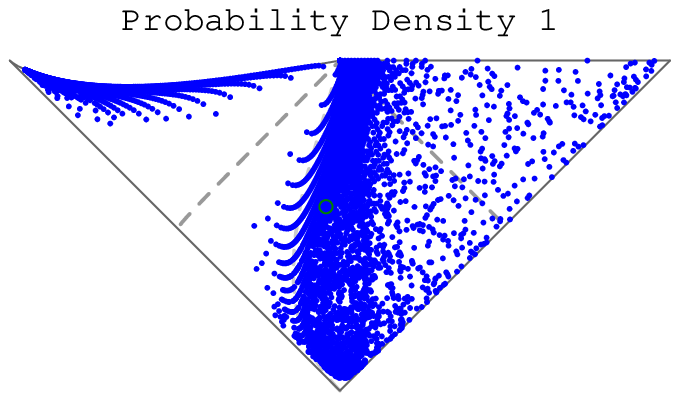}{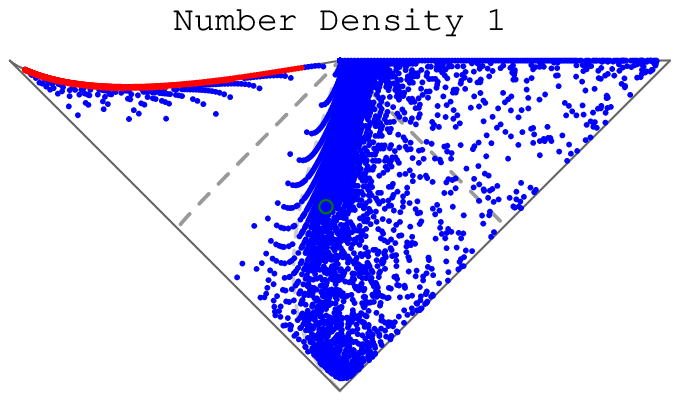}{Penrose density plots of probability
density and number density $\times$ mass of stationary mass 1}
The density plots show densities relative to a normalization point
in the center of the diagram just outside of the horizon indicated
by the small circle.  Once again, blue pixels indicate positive
values of that density, while red pixels indicate negative values. 
All probability densities are found to be positive semi-definite
throughout the global space-time.  One might note that as
previously stated, a positive semi-definite number density
requires negative mass solutions near the singularity, indicated
by the red values on the second density plot.

Scenarios involving various combinations of ingoing, outgoing,
and stationary massive quanta co-gravitating with ingoing and outgoing
massless quanta have been calculated.  The form of the four-velocities
for the given geometry are fixed by the particle type (e.g. massless vs.
massive).  The algebraic nature of
the quantum solutions make the numerical calculations straightforward.  To
obtain a given solution, one needs to adjust boundary values
for the phases and velocities in order to match the given physical boundary
conditions.  It is expected that one should be able to describe
physically mixed solutions (e.g. thermal systems or incoherent sums
of solutions consistent with the boundary conditions) 
in a direct manner using this formulation.

\setcounter{equation}{0}
\section{Conclusions}
\indent

The behaviors of classical particles and simple quantum
fields on a dynamic black hole have been both analytically
and numerically examined. 
The metric form taken to describe the dynamic black hole
directly corresponds with a Minkowski space-time asymptotically,
and utilizes the asymptotic observer's time coordinate to
describe the dynamics of the black hole without introducing
physically singular behavior at the horizon. 
This metric then defines the form for outgoing, ingoing, and in the
case of massive particles, stationary four-velocities.  These
four-velocity forms are then universally applicable on the
given geometry (principle of equivalence), and can be used to properly
incorporate geometrodynamics
(kinematics) without additional entanglements being introduced
to any (internal) quantum dynamics.
The trajectories of classical particles were plotted and found
to satisfy the constraints expected from causality on
the Penrose diagram.

In addition, simple quantum behaviors on the black hole
background were seen to follow intuitively consistent behaviors.
First, the form of the equation describing the dynamics of
spherically symmetric solutions to the massless Klein-Gordon
equation were developed.  Numerical solutions of
gravitating quanta on the
black hole background were found to 
have densities most prevalent
in the regions near the singularity and horizon.
The numerically explored
Klein-Gordon wavefunction was continuous throughout
the global space-time, but its derivative was found to behave
in a singular manner near the horizon
and the ingoing horizon.
The behavior of the energy-momentum density of the scalar
field was then compared with that necessary to generate
the dynamic black hole.  It was found that the
non-linear dependency of the Klein-Gordon
energy-momentum tensor
on field derivatives make it problematic for this scalar field to be
used to consistently mimic the radial dependency of
the Einstein tensor describing the dynamic black hole.

Next, generic complex scalar fields whose phase relationships are
linearly related to energy-momentum densities were developed. 
This means that the energy-momentum relationships that
generate coherent phase interference properties
(as experimentally observed) directly
relate to the energy-momentum tensor calculated for that field. 
The generic form can be generalized to spinor fields in a
straightforward manner.  The scalar form was shown to
satisfy the expected physical dynamics of quantum objects
(e.g. probability flux conservation, phase coherence, etc.) on the black
hole background.  Numerical solutions for the quantum state
vector describing the scalar fields depend only upon the boundary
conditions of any given quantum state, and are straightforward
to calculate.  The regions of the global space time available for
outgoing quanta are delineated by the black hole horizon, while
regions available for ingoing quanta are delineated by the
ingoing horizon of the black hole.  The energy-momentum
tensors of any of these quantum states linearly contribute to
the overall system kinematics and dynamics.

The construction of mutual/self gravitating macroscopic
collections of the scalar fields was next explored. 
A straightforward development of co-gravitating solutions
was expected to require the following properties:
(i) cluster decomposability of non-interacting radiations/quanta,
(ii)  proper time/affine dynamics for substantive kinematic flows
in order to incorporate the principles of relativity/equivalence, and
(iii)  linear forms for additivity of the energy-momentum contributions. 
The solutions found involved the collective
contributions of the energy-momentum densities
of the mutually gravitating quanta combined with a real,
spatially coherent \emph{core} gravitating field resulting in
a self-consistent gravitating system. 
The real core field for the excreting black hole examined consists of 
outgoing massless quanta 
which have a direct interaction term
with the metric form in the core Lagrangian.  
Because of the direct metric interaction, the core fields
do not have conserved probability densities, but have a form
which is fixed by that of the Einstein tensor.  The key
points of flexibility that allows generic solutions are that
the Euler-Lagrange equations for the core gravitating field
only require that the field be massless, while the co-gravitating
decomposable fields contribute to the overall energy-momentum
densities in an additive way directly related to the quantum
phase coherence energetics.

Since the approach here taken
is generic and not peculiar to the black
hole examined, it should be capable of describing several
physically relevant gravitational scenarios.   In particular,
this generic form should be
useful for developing mutually gravitating solutions for
static backgrounds in the exterior region
(e.g. planets, stars, etc.)
using ingoing and outgoing quantum
stationary states, along with
any  kinematically stationary solutions.
Fields (such as the generic gravitating quanta or gravitating clusters)
with vanishing extremal Lagrangian can always be algebraically
added to static geometries to help generate vacuum solutions to
Einstein's equation.

The author is involved with several
related on-going explorations of the
geometrodynamics of gravitating quanta.  One might gain insight into
the microscopic contributions of the individual quantum clusters
by more closely examining the single quantum limiting forms
of the macroscopic system (whether the system has
a singularity or not).  In addition,
the exploration of the co-gravitating initiation of
accretion to form a singularity
might answer fundamental questions on how
black holes start the process of collapse.
Finally, the construction of mutually gravitating cosmological
systems can be likewise accomplished in a straightforward
manner, as will be presented shortly in a follow up paper.

\begin{center}
\textbf{Acknowledgments}
\end{center}

The author gratefully acknowledges useful discussions with 
James Bjorken, Beth Brown, Tehani Finch, Tepper Gill, 
Harry Morrison, H. Pierre Noyes, Paul Sheldon, and Lenny Susskind.
Furthermore, the author wishes to dedicate this work to his
former student, dear friend, colleague, and co-researcher,
Beth A. Brown, who unexpectedly passed this month.


\begin{thebibliography}{99}

\bibitem{Overhauser}
R. Colella, A.W. Overhauser, and S.A. Werner, ``Observation of Gravitationally 
Induces Quantum Interference'', Phys.Rev.Lett. \textbf{34}, 1472-1474 
(1975). See also A.W.Overhauser and R.Colella, Phys.Rev.Lett. 
\textbf{33}, 1237 (1974). 

\bibitem{LMNP}
``Self-Consistent, Poincare-Invariant and Unitary 3-Particle Theory",
J. Lindesay, A. Markevich, H.P. Noyes, and G. Pastrana. 
Phys. Rev. \textbf{D 33}, 2339-2349 (1986).

\bibitem{Fadeev}
\textbf{Mathematical Aspects of the Three-Body Problem
in Quantum Scattering Theory}, L. D. Faddeev, (Davey, New York, 1965).

\bibitem{Dirac}
P.A.M. Dirac, Rev. Mod. Phys. \textbf{21}, 392 (1949). 
See also H. Leutwyler and J. Stern, Ann. Phys. \textbf{112},
94 (1978)

\bibitem{AKLN}
``A Non-Perturbative, Finite Particle Number Approach to Relativistic
Scattering Theory",  M.Alfred, P.Kwizera, J. Lindesay, and H.P.Noyes,
hep-th/0105241, Foundations of  Physics \textbf{34}(4), 581-616 (2004).

\bibitem{Compton}
``Construction of Non-Perturbative, Unitary Particle-Antiparticle Amplitudes
for Finite Particle Number Scattering Formalisms", J. Lindesay and H.P.Noyes,
Found. Phys. \textbf{35}(5), 39 pages (May 2005), DOI 10.1007/s10701-005-4563-8.
SLAC-PUB-9156, nucl-th/0203042, 47 pages (2002). 

\bibitem{Harry}
``The Geometry of Quantum Flow",  J. Lindesay and H. Morrison. 
\textbf{Mathematical Analysis of Physical Systems},
pp 135-167, ed. by R. Mickens. 
Van Nostrand Reinhold, Co., New York  (1985)

\bibitem{Gill}
``Canonical Proper Time Formulation for Physical Systems",
J. Lindesay and T. Gill, Foundations of Physics \textbf{34}(1), 169-182 (2004). 
See also ``Canonical Proper Time Formulation of Relativistic
Particle Dynamics", T. Gill and J. Lindesay,
International Journal of Theoretical Physics \textbf{32}, 2087-2098 (1993).

\bibitem{LinearSpinor}
``Linear Spinor Field Equations for Arbitrary Spins",
J. Lindesay, math-ph/0308003, 9 pages (2003).

\bibitem{rivermodel}
``The river model of black holes", A.J.S. Hamilton and
J.P. Lisle, gr-qc/0411060 (2004) 14 pages.

\bibitem{JLMay07}
``Coordinates with non-singular curvature for a time-dependent
black hole horizon",
J. Lindesay, gr-qc/0609019 (2006), \textit{Foundations of Physics} online 
DOI 10.1007/s10701-007-9146-4, 15 May 2007, 16 pages,.

\bibitem{BABJL1}
``Construction of a Penrose Diagram for a Spatially Coherent Evaporating Black Hole",
B. A. Brown and J. Lindesay, arXiv:0710.2032v1 [gr-qc] (2007) 12 pages.  
Class. Quantum Grav. \textbf{25} (2008) 105026 (8pp) doi:10.1088/0264-9381/25/10/105026.

\bibitem{BABJLphotons}
``Radial Photon Trajectories Near an Evaporating Black Hole Horizon,"
B. A. Brown and J. Lindesay, 
arXiv:0802.1660v1 [gr-qc] (2008) 9 pages.

\bibitem{LSJLBlackHoles}
\textbf{An Introduction to Black Holes, Information, 
and the String Theory Revolution: The Holographic Universe}, 
L. Susskind and J. Lindesay.   World Scientific, Singapore (Jan 2005), 
ISBN 981-256-083-1 hardback, ISBN 981-256-131-5 paperback.

\bibitem{JLNSBP08}
``An Exploration of the Physics of Spherically Symmetric Dynamic Horizons",
J. Lindesay
Invited talk presented at the Joint Annual Conference of the National
Society of Black Physicists and National Society of Hispanic Physicists
(20-24 February 2008), Washington, DC, 22 February 2008,
arXiv:0803.3018 [gr-qc] 25 pages.

\bibitem{Efimov}
V. Efimov, Phys. Let. \textbf{33B}, 563 (1970).  Also, V. Efimov,
Nucl. Phys. \textbf{A210}, 157 (1973).  For experimental evidence
see Hanns-Chistoph Nageri,
Nature \textbf{440}, 315 (2006)

\bibitem{WeinbergQFT}
\textbf{The Quantum Theory of Fields}, S. Weinberg, Cambridge
University Press, New York (1995) p. 66,
ISBN 0-521-55001-7.

\end{thebibliography}
\end{document}